\documentclass[12pt]{iopart}
\usepackage{epsfig}
\usepackage{iopams}
\begin{document}

\title[A Solution of the Hubbard Model]{A Solution of the Hubbard Model}

\author{Myung-Hoon Chung}

\address{College of Science and Technology, Hongik University,
Sejong 339-701, Korea}

\address{E-mail: mhchung@hongik.ac.kr}

\begin{abstract}
We report a ground-state solution for the two-dimensional
fermionic Hubbard model, which is obtained via a numerical
variational method. The two ingredients in this approach are
tensor network states and the time-evolving block decimation. We
easily handle the horizontal hopping in the Hamiltonian, and we
proceed further to observe the fermion-exchange effect caused by
the vertical hopping. By requiring no divergence and no
convergence to zero for the ground state, we successively
determine the ground-state energy per site as a function of the
chemical potential and the lattice length. In addition, we observe
saturation in the behavior of the ground-state energy as the
lattice length increases.
\end{abstract}




\maketitle

\section{Introduction}

In 1963, to understand the behavior of correlated electrons in
solids, a fermion lattice model was proposed independently by
three physicists: Martin Gutzwiller \cite{Gutzwiller}, Junjiro
Kanamori \cite{Kanamori}, and John Hubbard \cite{Hubbard1}. This
model has become widely known as the Hubbard model
\cite{Hubbard2}. Since this model's relevance to high $T_c$
superconductors was first suggested \cite{Dagotto}, much attention
has been paid to it. Recently, it has become possible to construct
experimental implementations of the Hubbard model using an optical
lattice for cold atoms \cite{Greiner,Jordens}, and hence, the
research community has refocused on the Hubbard model. Although
the model can be represented in a simple form, we encounter
notorious difficulties \cite{Troyer} when we attempt to find a
solution even numerically.

One of the main advances in the field of strongly correlated
systems is the establishment of the concept of the renormalization
group (RG) \cite{Wilson1}. In fact, Wilson also invented the
numerical RG (NRG) \cite{Wilson2} to solve the Kondo problem
\cite{Kondo}. Inspired by the NRG, White proposed the
density-matrix RG (DMRG) \cite{White}, which has proven to be a
great success in the simulation of strongly correlated
one-dimensional quantum lattice systems. It has been found that
the internal structure of the DMRG can be understood with respect
to the matrix-product states (MPS)
\cite{Ostlund,Garcia,Saberi,Schollwoeck}. For two-dimensional
systems, the projected entangled-pair states (PEPS)
\cite{Verstraete,Orus} are introduced. More generally, we call all
of these states tensor network states (TNS), and they include MPS,
PEPS, tree tensor network states \cite{Murg}, the multiscale
entanglement renormalization ansatz \cite{Vidal1}, and
matrix-product projected states \cite{Chou}. Beyond the spin-block
concept, the tensor network method based on the coarse-grained
tensor RG \cite{Levin} has been applied to a classical spin
system. The method was refined to the second RG \cite{Jiang,Xie}
by globally optimizing the truncation scheme and improving the
accuracy.

When a total Hamiltonian is written as a sum of local
Hamiltonians, Vidal \cite{Vidal2,Vidal3} introduced a powerful
method called time-evolving block decimation (TEBD) for finding
correlation functions. If the total Hamiltonian also has a type of
symmetry such as translational invariance, we can use the
so-called infinite TEBD \cite{Vidal4}, in which we assume that the
matrices in the TNS have the same form, and we update a few
matrices to achieve the ground state. However, because the TNS for
the Hubbard model is not an eigenstate of the number operator, the
TNS breaks the basic symmetry of particle-number preserving.
Furthermore, we do not insist on preserving the translational
invariance in the TNS. In consequence, we do not use the infinite
TEBD here. We alternatively adopt TEBD and extend it to the case
of the Hubbard model using PEPS. If the fermion-exchange effect is
involved during the TEBD procedure, a long-range entanglement
appears between the tensors of the PEPS. The essence of the
Hubbard model is to solve the problem caused by the
fermion-exchange effect.

In this paper, we focus on updating the large entangled part in
the TNS when we apply TEBD to the Hubbard model. To that end, we
first describe the nature of the TNS as an approximate ground
state for the Hubbard model. The connections between the tensors
in the TNS are represented by three types of bonds: horizontal,
vertical, and spin bonds. The set of the TNS is a small subspace
of the corresponding huge Hilbert space for the Hubbard model.
During the imaginary time evolution in TEBD, we restrict the
accessible states to the set of the TNS. Furthermore, in the
process of updating bonds, we adjust the proportional factor in
front of the state. By requiring no divergence and no convergence
to zero for the factor, we determine the form of the TNS for the
ground state and the corresponding energy.

This paper is organized as follows. In Sec. 2, a detailed
description of the Suzuki-Trotter decomposition is given, and we
introduce the tensor network state for the Hubbard model. In Sec.
3, using the Suzuki-Trotter decomposition, we present the
framework of the algorithm in the spirit of TEBD. Moreover, in
this section, we describe how to update the horizontal, vertical,
and spin bonds, and present the method of determining the
ground-state energy per site. In Sec. 4, we present consistency
checks for the method, and we summarize the numerical results
obtained when performing TEBD with small bond dimensions; the bond
dimensions should be increased in future works. The results for
the ground-state energy show evidence of saturation as the lattice
length increases, which indicates that the thermodynamic limit is
achieved. We observe the spin-flip symmetry breaking, and present
the critical strength of the on-site Coulomb repulsion.
In conclusion, we discuss a
parallelism for implementation in future work to improve the speed
of computing.

\section{Hamiltonian and Tensor Network States}

We begin by presenting the Hamiltonian for the Hubbard model,
which is written as
\begin{eqnarray}
H &=& -t\sum_{\langle i j
\rangle}(c^{\dagger}_{i\uparrow}c_{j\uparrow} +
c^{\dagger}_{j\uparrow}c_{i\uparrow} +
c^{\dagger}_{i\downarrow}c_{j\downarrow} +
c^{\dagger}_{j\downarrow}c_{i\downarrow}) \nonumber \\
& &
+U\sum_{i}(n_{i\uparrow}-\frac{1}{2})(n_{i\downarrow}-\frac{1}{2})-\mu
\sum_{i}(n_{i\uparrow}+n_{i\downarrow}) \nonumber \\
&=& H^{\uparrow}_{he} + H^{\uparrow}_{ho} + H^{\uparrow}_{ve} +
H^{\uparrow}_{vo} + H^{\downarrow}_{he} + H^{\downarrow}_{ho} +
H^{\downarrow}_{ve} + H^{\downarrow}_{vo} +  H_{d},
\end{eqnarray}
where $\langle i j \rangle$ represents nearest-neighbor hopping in
a two-dimensional lattice, and $n_{i\uparrow}$ and
$n_{i\downarrow}$ are the spin-up and the spin-down number
operators, respectively. We let the hopping strength $t$ be 1 and
vary the strengths of both the on-site Coulomb repulsion $U$ and
the chemical potential $\mu$; the number of fermions is controlled
by $\mu$. We divide the hopping term into four parts for each
spin, which are denoted by $h$(horizontal), $v$(vertical),
$e$(even), and $o$(odd), as shown in Fig. 1. The diagonal
Hamiltonian $H_{d}$ for a typical basis contains the last two
terms of the on-site repulsion and the chemical potential. The
Hubbard model may be the simplest quantum system of interacting
fermions on a lattice.

\begin{figure}
\includegraphics[width= 8 cm]{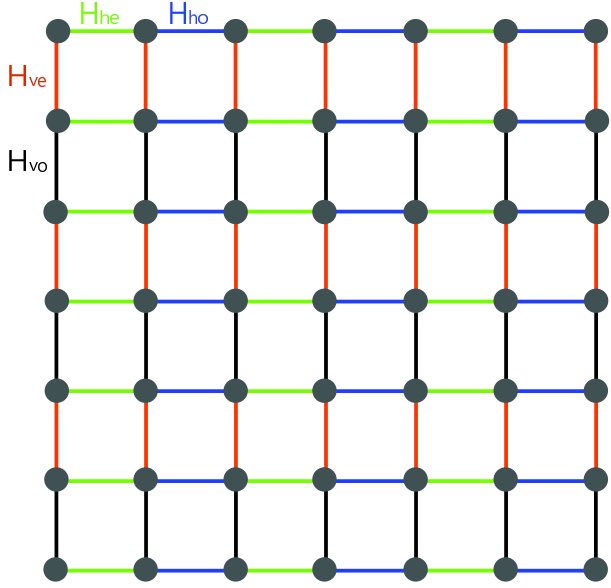}


\caption {(Color online) Connections between nearby points on the
square lattice. The connections are classified into four classes:
horizontal-even (denoted by $\langle i j \rangle_{he}$),
horizontal-odd, vertical-even, and vertical-odd. Here, the four
colors are used to represent four classes, and one hopping
Hamiltonian corresponds to each class.
 \label{fig:fig1} }
\end{figure}

We note that the Hamiltonian has symmetries. First of all, the
number operator $N_{op}=\sum_{i}(
c^{\dagger}_{i\uparrow}c_{i\uparrow} +
c^{\dagger}_{i\downarrow}c_{i\downarrow})$ commutes with the
Hamiltonian. When we impose the periodic boundary condition, the
translational symmetry appears. Furthermore, the Hamiltonian is
invariant under the spin-flip operation such as
$c_{i\uparrow}\rightarrow c_{i\downarrow}$ and
$c_{i\downarrow}\rightarrow c_{i\uparrow}$. We will discuss these
symmetries in relation to the TNS later.

For a given Hamiltonian $H$, we introduce an energy shift $E$ and
the inverse of the energy $T$, and then, we consider a formal
solution to the imaginary time Schr\"{o}dinger equation:
\begin{equation}
|\Psi( T ) \rangle = \exp\{ - ( H - E ) T \} |\Psi( 0 )\rangle.
\end{equation}
As $T$ goes to infinity, the state $|\Psi( T ) \rangle$ becomes
the ground state for properly chosen $E$. In fact, when $E$ is
larger or smaller than the ground-state energy, $|\Psi( T
)\rangle$ blows up or shrinks down, respectively, in the limit as
$T\rightarrow \infty$. In a numerical approach, we redefine $E$ as
a function of $T$ to determine the ground state.

We rewrite the operator using the Suzuki-Trotter decomposition
with a given small time step $\tau$ as
\begin{eqnarray}
\exp\{ - ( H - E ) T \}&\cong &
\prod^{T/\tau}\exp\{(E-H_{d})\tau\}
\nonumber \\
&\times&\Big[\mbox{The same expression
for spin down}\Big] \nonumber \\
&\times&\exp(-\frac{1}{4}H^{\uparrow}_{he}\tau)
\exp(-\frac{1}{2}H^{\uparrow}_{ho}\tau)
\exp(-\frac{1}{4}H^{\uparrow}_{he}\tau) \nonumber \\
&\times&\exp(-\frac{1}{2}H^{\uparrow}_{ve}\tau)
\exp(-H^{\uparrow}_{vo}\tau)
\exp(-\frac{1}{2}H^{\uparrow}_{ve}\tau) \nonumber \\
&\times&\exp(-\frac{1}{4}H^{\uparrow}_{he}\tau)
\exp(-\frac{1}{2}H^{\uparrow}_{ho}\tau)
\exp(-\frac{1}{4}H^{\uparrow}_{he}\tau).
\end{eqnarray}
It is not difficult to employ a higher-order Suzuki-Trotter
decomposition to obtain a more accurate calculation. Note that we
now decompose the operators in Eq. (3) in terms of elementary
operators such as
\begin{eqnarray}
\exp(-\frac{1}{4}H^{\uparrow}_{he}\tau) = \prod_{\langle i j
\rangle_{he}} \exp \{\frac{1}{4}t\tau
(c^{\dagger}_{i\uparrow}c_{j\uparrow} +
c^{\dagger}_{j\uparrow}c_{i\uparrow})\}, \nonumber \\
~~~~~~~~\vdots   \nonumber \\
\exp\{(E - H_{d})\tau \} = \prod_{i} \exp\{e\tau - U\tau
(n_{i\uparrow}-\frac{1}{2})(n_{i\downarrow}-\frac{1}{2}) +\mu \tau
(n_{i\uparrow}+n_{i\downarrow}) \}, \nonumber
\end{eqnarray}
where $e$ is the energy per site, that is, $E/N$, and
$N=\sum_{i}1$. Our strategy is to use Vidal's TEBD with these
elementary operators in a small subset of the Hilbert space. This
small subset is composed of the TNS characterized by the fixed
bond dimension.

For the fermionic Hubbard model, the usual tensor network states
should be suitably modified to describe fermions. In previous
works, many such attempts have been made; the Jordan-Wigner
strings was noticed in relation to fermions \cite{Barthel}, and we
find the fermionic projected entangled-pair states
\cite{Kraus,Corboz1,Pizorn,Corboz2,Corboz3} and the fermionic
multiscale entanglement renormalization ansatz
\cite{Pineda,Corboz4,Corboz5,Marti} for the ground states. These
fermionic modifications share some similarities, but they do not
agree with each other completely; thus, they require further
investigation. As a first step, we adopt the scheme of Corboz's
fermionic PEPS for our TNS, however for which we do not insist on
preserving the fermionic parity.

We use the one-to-one correspondence between a state of the
two-state chain and a state of the Fock space. The state of the
chain is represented by $\sigma_{i}$ and $\sigma_{N+i}$ for the
spin-up and the spin-down, respectively, and the state of the Fock
space is written in terms of the creation operators
$c^{\dagger}_{i\uparrow}$ and $c^{\dagger}_{i\downarrow}$ as
follows:
\begin{eqnarray}
&&| \sigma_{0} \cdots \sigma_{N-1}\sigma_{N} \cdots \sigma_{2N-1}
\rangle \nonumber \\
&&~~~~~~~~~~= (c^{\dagger}_{0\uparrow})^{\sigma_{0}} \cdots
(c^{\dagger}_{N-1\uparrow})^{\sigma_{N-1}}
(c^{\dagger}_{0\downarrow})^{\sigma_{N}} \cdots
(c^{\dagger}_{N-1\downarrow})^{\sigma_{2N-1}}|0 \rangle ,
\end{eqnarray}
where $\sigma_{N+i}= 0$ or $1$ means there is a spin-down fermion
vacancy or occupancy at the $i$-th site, respectively. It is
important to maintain the ordering of the fermions in the state of
the Fock space to handle the negative sign caused by the fermion
exchange. We adopt the zigzag ordering, which is the approach of
numbering sites from left to right and from right to left one by
one alternately in horizontal lines. For the example of $N=4
\times 4$, the corresponding ordering of sites is
\begin{equation*}
\begin{array}{ccccccccc}
0&1&2&3&~~~~~~&16&17&18&19 \\
7&6&5&4&~~~~~~&23&22&21&20 \\
8&9&10&11&~~~~~~&24&25&26&27 \\
15&14&13&12&~~~~~~&31&30&29&28
\end{array}
\end{equation*}
where the numbers from $0$ to $15$ denote the spin-up sites and
the numbers from $16$ to $31$ denote the spin-down sites.

When representing the TNS for the Hubbard model, we should
appreciate the area law for the entanglement entropy
\cite{Eisert}. Taking into account the square lattice, we include
two horizontal bonds and two vertical bonds for each tensor.
Because the TNS do not preserve the fermion numbers, it is natural
to break the translational symmetry also. Thus, we use different
tensors at all sites. In order to consider the general case of the
spin-flip symmetry breaking, we introduce different tensors for
the spin-down fermions from those for the spin-up fermions.
Because the Hubbard model has an on-site interaction, we connect
the two tensors of spin-up and spin-down at the same site using
the spin bond. In consequence, for each tensor we attach four legs
for the right, up, left, and down bonds and one leg for the spin
bond as well as the physical index. We assign a Schmidt
coefficient vector to each bond. Therefore, for the square system
of the length $L$, there are $2L^{2}$ tensors and $2L^{2} \times
2+L^{2}$ Schmidt coefficient vectors in our TNS as shown in Fig.
2. These tensors and vectors will be updated in the process of
TEBD with periodic boundary conditions.

A typical one of the $2L^{2}$ tensors, $A_{ruld}^{\sigma s}$, has
six indices, among which the physical index $\sigma$ takes a value
of $0$ or $1$. For the space-bond degree of freedom, the indices
$r$ (right), $u$ (up), $l$ (left), and $d$ (down) run from $0$ to
$\chi-1$, where $\chi$ is the bond dimension. For the spin-bond
degree of freedom, the index $s$ runs from $0$ to $\kappa -1$,
where $\kappa$ is the spin-bond dimension. A state in the space of
the tensor network states is written as
\begin{eqnarray}
|\mbox{TNS} \rangle =
\sum_{\cdots\sigma\rho\cdots\nu\eta\cdots\alpha \beta \cdots
\delta \gamma \cdots} &&\mbox{Tr} \left[
\begin{array}{cccc}
\ddots & \vdots & \vdots &  \\
\cdots & A^{\sigma}_{\uparrow}      & B^{\rho}_{\uparrow}    & \cdots \\
\cdots & C^{\eta}_{\uparrow}        & D^{\nu}_{\uparrow}     & \cdots \\
       & \vdots & \vdots &  \ddots \\
\end{array}
\right] \left[
\begin{array}{cccc}
\ddots & \vdots & \vdots &  \\
\cdots & A^{\alpha}_{\downarrow}      & B^{\beta}_{\downarrow}   & \cdots \\
\cdots & C^{\gamma}_{\downarrow}       & D^{\delta}_{\downarrow}    & \cdots \\
       & \vdots & \vdots &  \ddots \\
\end{array}
\right]  \nonumber \\
&&\times | \cdots \sigma \rho \cdots \nu \eta \cdots \cdots \alpha
\beta \cdots \delta \gamma \cdots\rangle ,
\end{eqnarray}
where we ignore the representation of the internal bond indices on
the tensors and all internal bonds are connected by $\mbox{Tr}$,
as shown in Fig. 2. Note that the zigzag ordering puts the
physical index $\nu$ before $\eta$ in the spin-like chain basis.
The thermodynamic limit will be achieved for $L \rightarrow
\infty$ and $\chi,\kappa \rightarrow \infty$.

\begin{figure}
\includegraphics[width= 14 cm]{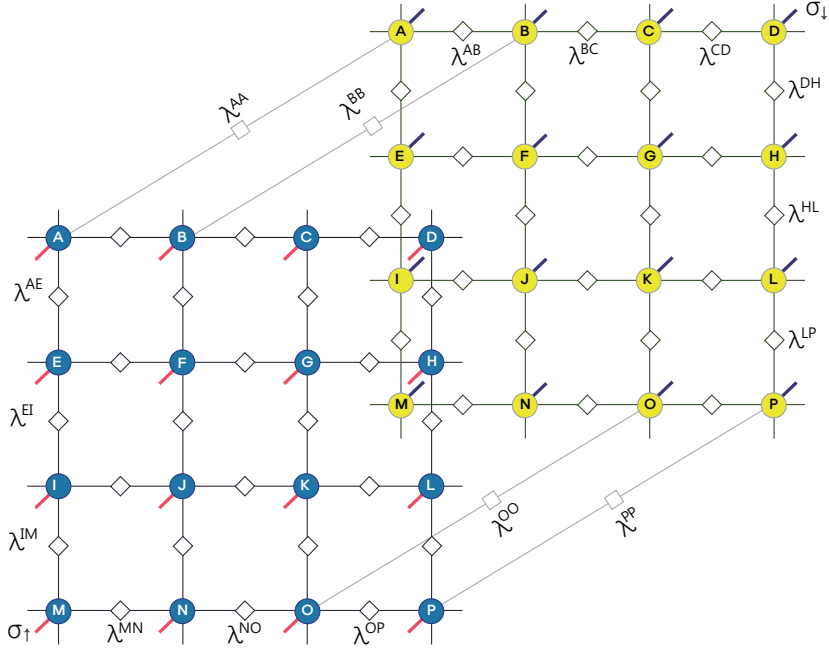}


\caption {A diagrammatic representation of a coefficient in front
of an orthonormal basis of $| \sigma_{0} \cdots
\sigma_{N-1}\sigma_{N} \cdots \sigma_{2N-1} \rangle $ for the case
of $N=L^{2}=4^{2}$. The closed circles represent $4^{2}\times 2$
six-index tensors. The open diamonds represent the Schmidt
coefficients $\lambda$ assigned to each bond. We neglect to draw
any spin bonds except the four lines between the spin-up and the
spin-down layers. Nevertheless, it should be understood that there
are spin bonds between all spin-up and spin-down tensors.
\label{fig:fig2} }
\end{figure}

When we consider the case of preserving the spin-flip symmetry, we
duplicate the tensors for spin-down using those for spin-up such
as
\begin{eqnarray}
|\mbox{TNS} \rangle =
\sum_{\cdots\sigma\rho\cdots\nu\eta\cdots\alpha \beta \cdots
\delta \gamma \cdots} &&\mbox{Tr} \left[
\begin{array}{cccc}
\ddots & \vdots & \vdots &  \\
\cdots & A^{\sigma}      & B^{\rho}    & \cdots \\
\cdots & C^{\eta}        & D^{\nu}     & \cdots \\
       & \vdots & \vdots &  \ddots \\
\end{array}
\right] \left[
\begin{array}{cccc}
\ddots & \vdots & \vdots &  \\
\cdots & A^{\alpha}    & B^{\beta}   & \cdots \\
\cdots & C^{\gamma}    & D^{\delta}  & \cdots \\
       & \vdots & \vdots &  \ddots \\
\end{array}
\right]  \nonumber \\
&&\times | \cdots \sigma \rho \cdots \nu \eta \cdots \cdots \alpha
\beta \cdots \delta \gamma \cdots\rangle .
\end{eqnarray}
In the process of TEBD, we update $L^{2}$ tensors and $L^{2}
\times (2+1)$ Schmidt coefficient vectors for the ground state
preserving the spin-flip symmetry.

The operator $\exp\{(E - H) \tau \}$ of Eq. (3) acts on the state
$|\mbox{TNS}\rangle$ of Eq. (5) consecutively. Thus, the output
state $\exp\{(E - H) \tau \}|\mbox{TNS}\rangle$, which is outside
the space of the TNS, is approximated into a TNS by updating the
tensors and the vectors. The updating procedure is the subject of
the next section.

\section{The Updating Procedure}

In order to proceed with TEBD, we consider the single elementary
hopping term $\exp\{t\tau (c^{\dagger}_{i\uparrow}c_{j\uparrow} +
c^{\dagger}_{j\uparrow}c_{i\uparrow})\}$, where we add a factor of
$1/2$ or $1/4$ in front of $t\tau$ if necessary. When the
elementary operator acts on the previous TNS, we approximate the
output state into our subset of the TNS by updating the tensors
and the vectors locally. This procedure is the basic strategy of
TEBD.

When the elementary operator $\exp\{t\tau
(c^{\dagger}_{i\uparrow}c_{j\uparrow} +
c^{\dagger}_{j\uparrow}c_{i\uparrow})\}$ acts on a basis vector $|
\sigma_{0}\cdots \sigma_{i} \cdots \sigma_{j} \cdots
\sigma_{N-1}\sigma_{N}\cdots\sigma_{2N-1} \rangle$, we find the
following important result \cite{Chung0}, which is written for
four cases that correspond to $\sigma_i = 0$ or $1$ and $\sigma_j
= 0$ or $1$:
\begin{equation}
\left\{
\begin{array}{l}
\exp\{t\tau (c^{\dagger}_{i\uparrow}c_{j\uparrow} +
c^{\dagger}_{j\uparrow}c_{i\uparrow})\} | \cdots 0 \cdots 0 \cdots
\rangle = |
\cdots 0 \cdots 0 \cdots \rangle \nonumber\\
\exp\{t\tau (c^{\dagger}_{i\uparrow}c_{j\uparrow} +
c^{\dagger}_{j\uparrow}c_{i\uparrow})\} | \cdots 0 \cdots 1 \cdots
\rangle = \cosh (t\tau) |
\cdots 0 \cdots 1 \cdots \rangle \nonumber\\
~~~~~~~~~~~~~~~~~~~~~~~~~~~~ +\sinh
(t\tau)(-1)^{\sigma_{i+1}+\cdots+\sigma_{j-1}}
| \cdots 1 \cdots 0 \cdots \rangle \nonumber\\
\exp\{t\tau (c^{\dagger}_{i\uparrow}c_{j\uparrow} +
c^{\dagger}_{j\uparrow}c_{i\uparrow})\} | \cdots 1 \cdots 0 \cdots
\rangle = \cosh (t\tau) |
\cdots 1 \cdots 0 \cdots \rangle \nonumber\\
~~~~~~~~~~~~~~~~~~~~~~~~~~~~ +\sinh
(t\tau)(-1)^{\sigma_{i+1}+\cdots+\sigma_{j-1}}
| \cdots 0 \cdots 1 \cdots \rangle \nonumber\\
\exp\{t\tau (c^{\dagger}_{i\uparrow}c_{j\uparrow} +
c^{\dagger}_{j\uparrow}c_{i\uparrow})\} | \cdots 1 \cdots 1 \cdots
\rangle = | \cdots 1 \cdots 1 \cdots \rangle \nonumber
\end{array} \right .
\end{equation}
The sign of $(-1)^{\sigma_{i+1}+\cdots+\sigma_{j-1}}$ reflects the
fermion-exchange effect. Note that the values of the physical
indices at the sites numbered from $i+1$ to $j-1$ are related to
the sign, which makes it difficult to handle the vertical hopping.
The above equations play a key role in updating the TNS.

\subsection{Updates to the Horizontal Bonds}

Because a horizontal bond connects a site to the next site, that
is, to $j=i+1$ in our ordering of sites, it is straightforward to
update horizontal bonds. As we can see in the Suzuki-Trotter
decomposition of Eq. (3), we first handle $H_{he}$, then $H_{ho}$,
and then $H_{he}$ again. Here, we present the typical procedure
for updating a horizontal bond.

For example, to update $A$, $B$, and $\lambda^{AB}$ in Fig. 2 with
the periodic boundary condition, we consider the tensor product
that is represented symbolically as follows:
\begin{equation}
(A \otimes B )^{\sigma s \rho
\tilde{s}}_{uld\tilde{r}\tilde{u}\tilde{d}} \equiv
\sum_{x=0}^{\chi -1} A^{\sigma s}_{xuld} \lambda^{AB}_x B^{\rho
\tilde{s}}_{\tilde{r}\tilde{u}x\tilde{d}} \lambda^{AA}_s
\lambda^{MA}_u \lambda^{DA}_l \lambda^{AE}_d
\lambda^{BB}_{\tilde{s}}
\lambda^{BC}_{\tilde{r}}\lambda^{NB}_{\tilde{u}}\lambda^{BF}_{\tilde{d}},
\end{equation}
where the eight Schmidt coefficients are attached to $A$ and $B$.
Using the result of Eq. (7), we find the ten-index tensor
$\Theta^{\sigma s \rho\tilde{s}}_{uld\tilde{r}\tilde{u}\tilde{d}}$
to update $A$, $B$, and $\lambda^{AB}$:
\begin{equation}
\left\{
\begin{array}{l}
\Theta^{0s0\tilde{s}}_{uld\tilde{r}\tilde{u}\tilde{d}} = (A
\otimes B )^{0 s 0
\tilde{s}}_{uld\tilde{r}\tilde{u}\tilde{d}} \nonumber\\
\Theta^{0s1\tilde{s}}_{uld\tilde{r}\tilde{u}\tilde{d}} = \cosh
(t\tau)\times(A \otimes B )^{0 s 1
\tilde{s}}_{uld\tilde{r}\tilde{u}\tilde{d}}
 +  \sinh (t\tau)\times(A \otimes B )^{1 s 0
\tilde{s}}_{uld\tilde{r}\tilde{u}\tilde{d}}  \nonumber\\
\Theta^{1s0\tilde{s}}_{uld\tilde{r}\tilde{u}\tilde{d}} = \cosh
(t\tau)\times(A \otimes B )^{1 s 0
\tilde{s}}_{uld\tilde{r}\tilde{u}\tilde{d}}
 + \sinh (t\tau)\times (A \otimes B )^{0 s 1
\tilde{s}}_{uld\tilde{r}\tilde{u}\tilde{d}}  \nonumber\\
\Theta^{1s1\tilde{s}}_{uld\tilde{r}\tilde{u}\tilde{d}}=(A \otimes
B )^{1 s 1 \tilde{s}}_{uld\tilde{r}\tilde{u}\tilde{d}} \nonumber
\end{array} \right .
\end{equation}
We emphasize the physical-index exchange between $0$ and $1$ in
the tensor product multiplied by $\sinh(t\tau)$. By employing
singular value decompositions (SVD), we obtain the updated
$\tilde{\lambda}^{AB}_{x}$ by keeping the $\chi$ largest weights:
\begin{eqnarray}
\Theta^{\sigma s \rho \tilde{s}}_{uld\tilde{r}\tilde{u}\tilde{d}}
&\rightarrow& \sum_{x=0}^{\chi-1} \bar{A}^{\sigma s}_{xuld}
\tilde{\lambda}^{AB}_x \bar{B}^{\rho
\tilde{s}}_{\tilde{r}\tilde{u}x\tilde{d}}\nonumber \\
&=&\sum_{x=0}^{\chi-1} \tilde{A}^{\sigma s}_{xuld}
\tilde{\lambda}^{AB}_x \tilde{B}^{\rho
\tilde{s}}_{\tilde{r}\tilde{u}x\tilde{d}}\lambda^{AA}_s
\lambda^{MA}_u \lambda^{DA}_l \lambda^{AE}_d
\lambda^{BB}_{\tilde{s}}
\lambda^{BC}_{\tilde{r}}\lambda^{NB}_{\tilde{u}}\lambda^{BF}_{\tilde{d}}.
\end{eqnarray}
By dividing and attaching the eight weights, we find $\tilde{A}$
and $\tilde{B}$ in the above. We denote this process graphically
as follows:
\begin{equation*}
\begin{array}{ccccccccccc}
 &|&        &|& &                   & &        |& &        |&  \\
-& &\Theta  & &-&     \rightarrow   &-&\tilde{A}&-&\tilde{B}&- \\
 &|&        &|& &                   & &        |& &        |&
\end{array}
\end{equation*}
where we omit both the spin bonds and the physical indices.

A similar procedure is performed for other tensors and other
vectors. By updating all $2L^{2}$ tensors, we finish the
horizontal-bond update. We note that it is possible to update all
$2L^{2}$ tensors simultaneously if we use multi-core computers.
Thus, we can easily parallelize the horizontal-bond update.

\subsection{Updates to the Vertical Bonds}

The vertical bonds exhibit a striking difference from the
horizontal bonds during the update process: the notorious
fermion-exchange effect appears when fermions are hopping
vertically. For the vertical bonds, we should consider
$\exp\{t\tau (c^{\dagger}_{i\uparrow}c_{j\uparrow} +
c^{\dagger}_{j\uparrow}c_{i\uparrow})\}|\cdots\sigma_i
\cdots\sigma_j \cdots \rangle$, where the number of sites between
$i$ and $j$ is given by a value from $0$ to $2L-2$ in the zigzag
ordering. Therefore, all tensors at the sites between $i$ and $j$
should be updated. Here, we introduce the method for updating the
tensors between $i$ and $j$ one by one.

For example, to update $A$, $E$, and $\lambda^{AE}$ on the
vertical bond in Fig. 2 with the periodic boundary condition, we
begin by writing the tensor product of $A$ and $E$ as follows:
\begin{equation}
\left(
\begin{array}{l}
A \\
\otimes \\
E
\end{array}
\right){}^{\sigma s \eta
\tilde{s}}_{rul\tilde{r}\tilde{l}\tilde{d}} \equiv
\sum_{x=0}^{\chi-1} A^{\sigma s}_{rulx} \lambda^{AE}_x E^{\eta
\tilde{s}}_{\tilde{r}x\tilde{l}\tilde{d}} \lambda^{AA}_s
\lambda^{AB}_r \lambda^{MA}_u \lambda^{DA}_l
\lambda^{EE}_{\tilde{s}}
\lambda^{EF}_{\tilde{r}}\lambda^{HE}_{\tilde{l}}\lambda^{EI}_{\tilde{d}}.
\end{equation}
From the result of Eq. (7), when the vertical-hopping term acts on
the TNS we find the portion that should be updated into a single
tensor network:
\begin{equation*}
\begin{array}{cccccccccccc}
 & & |       & &|& &|& &|&  \\
 &-&         &-&B&-&C&-&D&- \\
 & & \Phi    & &|& &|& &|&  \\
 &-&         &-&F&-&G&-&H&- \\
 & & |       & &|& &|& &|&  \\  \\
 & & |       & &|& &|& &|&  \\
 &-&         &-&(-1)^{\rho}B&-&(-1)^{\alpha} C&-&(-1)^{\gamma} D&- \\
+& &\Psi     & &           |& &              |& &              |&  \\
 &-&         &-&(-1)^{\nu} F&-&(-1)^{\beta}  G&-&(-1)^{\delta} H&- \\
 & & |       & &|& &|& &|&  \\
\end{array}
\end{equation*}
where we omit the spin bonds and the legs for the physical
indices. Because of the fermion exchange, the many signs appear in
front of the tensors in the second term. Each power of $(-1)$,
such as $\rho$, $\nu$, $\alpha$, $\beta$, $\gamma$, or $\delta$,
is the physical index of the corresponding tensor. Just as we
introduce a ten-index tensor for the horizontal-bond update, we
similarly find two ten-index tensors for the vertical-bond update,
namely, $\Phi^{\sigma s \eta
\tilde{s}}_{rul\tilde{r}\tilde{l}\tilde{d}}$ and $\Psi^{\sigma s
\eta \tilde{s}}_{rul\tilde{r}\tilde{l}\tilde{d}}$, which are
written in terms of the tensor product as follows:
\begin{equation}
\begin{array}{ll}
\Phi^{0 s 0 \tilde{s}}_{rul\tilde{r}\tilde{l}\tilde{d}} = \left(
\begin{array}{l}
A \\
\otimes \\
E
\end{array}
\right){}^{0 s 0 \tilde{s}}_{rul\tilde{r}\tilde{l}\tilde{d}} &
\Psi^{0 s 0 \tilde{s}}_{rul\tilde{r}\tilde{l}\tilde{d}}= 0
\\
\Phi^{0 s 1 \tilde{s}}_{rul\tilde{r}\tilde{l}\tilde{d}} =
\cosh(t\tau)\times\left(
\begin{array}{l}
A \\
\otimes \\
E
\end{array}
\right){}^{0 s 1 \tilde{s}}_{rul\tilde{r}\tilde{l}\tilde{d}} &
\Psi^{0 s 1
\tilde{s}}_{rul\tilde{r}\tilde{l}\tilde{d}}=\sinh(t\tau)\times\left(
\begin{array}{l}
A \\
\otimes \\
E
\end{array}
\right){}^{1 s 0 \tilde{s}}_{rul\tilde{r}\tilde{l}\tilde{d}}
\\
\Phi^{1 s 0 \tilde{s}}_{rul\tilde{r}\tilde{l}\tilde{d}} =
\cosh(t\tau)\times\left(
\begin{array}{l}
A \\
\otimes \\
E
\end{array}
\right){}^{1 s 0 \tilde{s}}_{rul\tilde{r}\tilde{l}\tilde{d}} &
\Psi^{1 s 0
\tilde{s}}_{rul\tilde{r}\tilde{l}\tilde{d}}=\sinh(t\tau)\times\left(
\begin{array}{l}
A \\
\otimes \\
E
\end{array}
\right){}^{0 s 1 \tilde{s}}_{rul\tilde{r}\tilde{l}\tilde{d}}
\\
\Phi^{1 s 1 \tilde{s}}_{rul\tilde{r}\tilde{l}\tilde{d}} = \left(
\begin{array}{l}
A \\
\otimes \\
E
\end{array}
\right){}^{1 s 1 \tilde{s}}_{rul\tilde{r}\tilde{l}\tilde{d}} &
\Psi^{1 s 1 \tilde{s}}_{rul\tilde{r}\tilde{l}\tilde{d}}= 0
\end{array}
\end{equation}

At this point, we propose a crucial idea to update the long tensor
chain given above. We call this idea {\it doubling}. Doubling
means that we enlarge the bond dimension for the indices
$r$(right) and $l$(left) by a factor of two such that they now run
from $0$ to $2\chi -1$. Graphically, doubling is represented by
changing from $-B-$ to $=B=$, and similarly for $=C=$ and other
tensors. Explicitly, we let
\begin{equation}
(=B=)^{\rho s}_{ruld} \equiv \left\{
\begin{array}{ll}
B^{\rho s}_{ruld}&~~\mbox{for}~~r < \chi~\mbox{and}~l<\chi \nonumber\\
(-1)^{\rho}B^{\rho s}_{(r-\chi)u(l-\chi)d}&~~\mbox{for}~~r \ge \chi~\mbox{and}~l \ge \chi \nonumber\\
0&~~\mbox{otherwise} \nonumber
\end{array} \right .
\end{equation}
Correspondingly, the ten-index tensors $\Phi$ and $\Psi$ are
combined into $\Theta$ as follows:
\begin{equation}
({}^{-}_{-}\Theta^{=}_{=})^{\sigma s
\eta\tilde{s}}_{rul\tilde{r}\tilde{l}\tilde{d}} \equiv \left\{
\begin{array}{ll}
\Phi^{\sigma s
\eta\tilde{s}}_{rul\tilde{r}\tilde{l}\tilde{d}}&~~\mbox{for}~~r < \chi~\mbox{and}~\tilde{r}<\chi \nonumber\\
\Psi^{\sigma s
\eta\tilde{s}}_{(r-\chi)ul(\tilde{r}-\chi)\tilde{l}\tilde{d}}&~~\mbox{for}~~r \ge \chi~\mbox{and}~\tilde{r} \ge \chi \nonumber\\
0&~~\mbox{otherwise} \nonumber
\end{array} \right .
\end{equation}
Obviously, the vectors with the enlarged bond dimensions are
defined as
\begin{equation}
\lambda^{AB}_{r} \equiv \left\{
\begin{array}{ll}
\lambda^{AB}_{r}&~~\mbox{for}~~r < \chi     \nonumber\\
\lambda^{AB}_{r-\chi}&~~\mbox{for}~~r \ge \chi \nonumber
\end{array} \right .
\end{equation}
As a result of doubling, the addition of two tensor networks can
be written as a single tensor network but with the increased bond
dimensions for the $r$ and $l$ indices. In consequence, we can
write the chain as
\begin{equation*}
\begin{array}{cccccccccccc}
 & & |       & &|& &|& &|&  \\
 &-&         &=&B&=&C&=&D&- \\
 & & \Theta  & &|& &|& &\parallel&  \\
 &-&         &=&F&=&G&=&H&- \\
 & & |       & &|& &|& &|&
\end{array}
\end{equation*}
where the rightmost tensors $D$ and $H$ have the doubled indices
$d$ and $u$, respectively, and are as follows:
\begin{equation}
(=D-)^{\gamma s}_{ruld} \equiv \left\{
\begin{array}{ll}
D^{\gamma s}_{ruld}&~~\mbox{for}~~d < \chi~\mbox{and}~l<\chi \nonumber\\
(-1)^{\gamma}D^{\gamma s}_{ru(l-\chi)(d-\chi)}&~~\mbox{for}~~d \ge \chi~\mbox{and}~l \ge \chi \nonumber\\
0&~~\mbox{otherwise} \nonumber
\end{array} \right .
\end{equation}
\begin{equation}
(=H-)^{\delta s}_{ruld} \equiv \left\{
\begin{array}{ll}
H^{\delta s}_{ruld}&~~\mbox{for}~~u < \chi~\mbox{and}~l<\chi \nonumber\\
(-1)^{\delta}H^{\delta s}_{r(u-\chi)(l-\chi)d}&~~\mbox{for}~~u \ge \chi~\mbox{and}~l \ge \chi \nonumber\\
0&~~\mbox{otherwise} \nonumber
\end{array} \right .
\end{equation}
It is useful to see the matrix forms of $\Theta$ and $B$ written
in the following way:
\begin{equation}
{}^{-}_{-}\Theta^{=}_{=} \equiv \left(
\begin{array}{c|c}
{}^{-}_{-}\Phi^{-}_{-} & 0 \\
\hline 0 & {}^{-}_{-}\Psi^{-}_{-}
\end{array}
\right) ~~~\mbox{and}~~~  =B= \equiv \left(
\begin{array}{c|c}
-B- & 0 \\
\hline 0 & -(-1)^{\rho}B-
\end{array}
\right)
\end{equation}
and the similar forms for $D$ and $H$.

To maintain the bond dimension, we make an approximation using
SVD. We perform SVD for $\Theta$ first; then, we obtain
$\acute{A}$, $\acute{E}$, and simultaneously, we obtain the vector
$\tilde{\lambda}^{AE}$. Next, we perform SVD again from
$-\acute{A}=B=$ to $-\tilde{A}-\acute{B}=$ as follows:
\begin{equation*}
\begin{array}{cccccccccccccccccc}
 & |       & &|& &            & &       | & &|& &            & &       | & &       | &  \\
-&         &=&B&=&            &-&\acute{A}&=&B&=&            &-&\tilde{A}&-&\acute{B}&= \\
 & \Theta  & &|& &\rightarrow & &       | & &|& &\rightarrow & &       | & &       | &  \\
-&         &=&F&=&            &-&\acute{E}&=&F&=&            &-&\tilde{E}&-&\acute{F}&= \\
 & |       & &|& &            & &       | & &|& &            & &       | & &       | &
\end{array}
\end{equation*}
For $B$ and $C$ in Fig. 2, we change from $-\acute{B}=C=$ to
$-\tilde{B}-\acute{C}=$ by using SVD. We continue performing SVD
tensor by tensor until we reach the rightmost $D$. We also do the
same thing for the lower half-chain from $E$ to $H$. Finally, we
obtain $\acute{D}$ and $\acute{H}$, and we perform SVD as follows:
\begin{equation*}
\begin{array}{ccccccccc}
 &        |& &                   & &        |&   \\
-&\acute{D}&-&                   &-&\tilde{D}&-  \\
 &\parallel& &    \rightarrow    & &        |&   \\
-&\acute{H}&-&                   &-&\tilde{H}&-  \\
 &        |& &                   & &        |&
\end{array}
\end{equation*}
where the right-hand bonds of $D$ and $H$ are connected to the
left-hand bonds of the leftmost tensors by the periodic boundary
condition.

It is worth noting that we can approximate the tensor chain in
different orderings. For example, we perform SVD first for the
bond between $B$ and $C$ or $A$ and $B$ as shown below:
\begin{equation*}
\begin{array}{cccccccccc}
           &-&A&=&B&=&C&=&D&-\\
\rightarrow&-&A&=&B&-&C&=&D&-\\
\rightarrow&-&A&-&B&-&C&=&D&-\\
\rightarrow&-&A&-&B&-&C&-&D&-
\end{array}
\end{equation*}
or
\begin{equation*}
\begin{array}{cccccccccc}
           &-&A&=&B&=&C&=&D&-\\
\rightarrow&-&A&-&B&=&C&=&D&-\\
\rightarrow&-&A&-&B&-&C&=&D&-\\
\rightarrow&-&A&-&B&-&C&-&D&-
\end{array}
\end{equation*}
We note that there are twice as many singular values for
$=B=C=\rightarrow =B-C=$ as those for $-B=C=\rightarrow -B-C=$ in
the approximation by SVD. Because we keep only a fixed number of
singular values, we lose more for $=B=C=\rightarrow =B-C=$ than
for $-B=C=\rightarrow -B-C=$. It is reasonable to perform SVD one
by one from the end as our scheme above.

When we apply $\exp(-\frac{1}{2}H_{ve}^{\uparrow}\tau)$ to the
tensor network state, we assume that the elementary operators act
on the tensors one by one from right to left. Thus, the vectors on
the horizontal bonds are updated repeatedly. This convention is
different from the case of the horizontal hopping
$\exp(-\frac{1}{4}H_{he}^{\uparrow}\tau)$, which updates the
vectors on the horizontal bonds only once. As a result, after we
perform SVD repeatedly for
$\exp(-\frac{1}{2}H_{ve}^{\uparrow}\tau)$, we return to the same
form of the tensor network state with modified tensors and
vectors:
\begin{equation*}
\begin{array}{ccccccccccccccccc}
 &        |& &        |& &        |& &        |& \\
-&\tilde{A}&-&\tilde{B}&-&\tilde{C}&-&\tilde{D}&-\\
 &        |& &        |& &        |& &        |& \\
-&\tilde{E}&-&\tilde{F}&-&\tilde{G}&-&\tilde{H}&-\\
 &        |& &        |& &        |& &        |&
\end{array}
\end{equation*}

For $\exp(-H_{vo}^{\uparrow}\tau)$, we follow a similar procedure
for the odd vertical bonds for example between $H$ and $L$ in Fig.
2. After doubling, we represent the portion that should be updated
as
\begin{equation*}
\begin{array}{ccccccccccc}
 &|         & &|& &|& &      |&  \\
-&E         &=&F&=&G&=&       &- \\
 &\parallel & &|& &|& & \Theta&  \\
-&I         &=&J&=&K&=&       &- \\
 &|         & &|& &|& &      |&  \\
\end{array}
\end{equation*}
We assume that the elementary operators in
$\exp(-H_{vo}^{\uparrow}\tau)$ act from left to right. Because
doubling should be performed in the left-hand part of our zigzag
ordering, $\Theta$ is in the right-hand part of this network. We
follow the same procedure for approximation: via the SVD of
${}^{=}_{=}\Theta^{-}_{-}$, we obtain the tensors of $=\acute{H}-$
and $=\acute{L}-$, and simultaneously, we obtain the updated
vector $\tilde{\lambda}^{HL}$. Again, we repeat the SVD process to
reduce the doubled bond dimensions to the original dimensions. In
the end, we obtain updated tensors and vectors.

\subsection{Updates to the Spin Bonds}

Because the elementary operator for the spin bond $\exp\{e\tau -
U\tau (n_{i\uparrow}-\frac{1}{2})(n_{i\downarrow}-\frac{1}{2})+\mu
\tau (n_{i\uparrow}+n_{i\downarrow}) \}$ is diagonal with respect
to our typical base vectors, we obtain
\begin{eqnarray}
&&\exp\{e\tau - U\tau
(n_{i\uparrow}-\frac{1}{2})(n_{i\downarrow}-\frac{1}{2})+\mu \tau
(n_{i\uparrow}+n_{i\downarrow}) \}|\cdots\sigma_i
\cdots\sigma_{N+i}\cdots \rangle
\nonumber \\
&&~= \exp\{e\tau -U\tau(\sigma_i
-\frac{1}{2})(\sigma_{N+i}-\frac{1}{2}) + \mu\tau (\sigma_i +
\sigma_{N+i} ) \}|\cdots\sigma_i \cdots\sigma_{N+i}\cdots \rangle
. \nonumber
\end{eqnarray}
From this equation, we determine the ten-index tensor
$\Theta^{\sigma
\tilde{\sigma}}_{ruld\tilde{r}\tilde{u}\tilde{l}\tilde{d}}$ to
update the vectors on the spin bonds as follows:
\begin{equation}
\Theta^{\sigma
\tilde{\sigma}}_{ruld\tilde{r}\tilde{u}\tilde{l}\tilde{d}}=
\exp\{e\tau -U\tau(\sigma-\frac{1}{2})(\tilde{\sigma}-\frac{1}{2})
+ \mu\tau (\sigma + \tilde{\sigma}) \} \times \left[
\begin{array}{c}
A_{\uparrow} \\
\otimes \\
A_{\downarrow}
\end{array}
\right]{}^{\sigma
\tilde{\sigma}}_{ruld\tilde{r}\tilde{u}\tilde{l}\tilde{d}}
\end{equation}
where the tensor product is given by
\begin{equation}
\left[
\begin{array}{c}
A_{\uparrow} \\
\otimes \\
A_{\downarrow}
\end{array}
\right]{}^{\sigma
\tilde{\sigma}}_{ruld\tilde{r}\tilde{u}\tilde{l}\tilde{d}} \equiv
\sum_{x=0}^{\kappa -1} A^{\sigma x}_{\uparrow ruld} \lambda^{AA}_x
A^{\tilde{\sigma} x}_{\downarrow
\tilde{r}\tilde{u}\tilde{l}\tilde{d}} \lambda^{AB}_{\uparrow r}
\lambda^{MA}_{\uparrow u} \lambda^{DA}_{\uparrow l}
\lambda^{AE}_{\uparrow d} \lambda^{AB}_{\downarrow\tilde{r}}
\lambda^{MA}_{\downarrow\tilde{u}}\lambda^{DA}_{\downarrow\tilde{l}}\lambda^{AE}_{\downarrow\tilde{d}}
\end{equation}
with the periodic boundary condition in Fig. 2.

As in the horizontal-bond update, we perform SVD for $\Theta$ to
find $\tilde{\lambda}^{AA}$. By dividing and attaching the eight
vectors, we obtain the tensor update $\tilde{A}$:
\begin{eqnarray}
\Theta^{\sigma \tilde{\sigma}
}_{ruld\tilde{r}\tilde{u}\tilde{l}\tilde{d}} &\rightarrow&
\sum_{x=0}^{\kappa-1} \bar{A}^{\sigma x}_{\uparrow ruld}
\tilde{\lambda}^{AA}_x \bar{A}^{\tilde{\sigma}
x}_{\downarrow\tilde{r}\tilde{u}\tilde{l}\tilde{d}}\nonumber \\
&=&\sum_{x=0}^{\kappa -1} \tilde{A}^{\sigma x}_{\uparrow ruld}
\tilde{\lambda}^{AA}_x \tilde{A}^{\tilde{\sigma} x}_{\downarrow
\tilde{r}\tilde{u}\tilde{l}\tilde{d}} \lambda^{AB}_{\uparrow r}
\lambda^{MA}_{\uparrow u} \lambda^{DA}_{\uparrow l}
\lambda^{AE}_{\uparrow d} \lambda^{AB}_{\downarrow\tilde{r}}
\lambda^{MA}_{\downarrow\tilde{u}}\lambda^{DA}_{\downarrow\tilde{l}}\lambda^{AE}_{\downarrow\tilde{d}}.
\end{eqnarray}

We follow the same procedure for all $2L^{2}$ tensors. It is easy
to parallelize this process using multi-core computers.

\subsection{Energy Updates}

Being inspired by the diffusion Monte Carlo
\cite{Ceperley,Chung1}, we introduced the energy per site $e$ in
the operator of the Suzuki-Trotter decomposition. While the energy
in the diffusion Monte Carlo is adjusted by controlling the number
of replicas, here we determine $e$ by managing the factor in front
of the wave function. The algorithm is as follows: when a typical
operator $\exp(h\tau)$ acts on a tensor network state $|\mbox{TNS}
\rangle$, we perform SVD and obtain $\chi$ singular values of
$\lambda_{0} \ge \lambda_{1} \ge \cdots \ge \lambda_{\chi -1}$. We
take $\lambda_{0}$ and place it in front of the wave function, and
we modify the singular values as follows: $1 \ge
\lambda_{1}/\lambda_{0} \ge \cdots \ge \lambda_{\chi
-1}/\lambda_{0}$. In this way, we normalize $|\mbox{TNS} \rangle$
such that all $5L^{2}$ vectors on each bond have the maximum value
of 1. Thus, whenever the weights are modified by
$\exp(E_{k}\tau-H\tau)$ acting on the $k$-th time step state
$|\mbox{TNS}_{k} \rangle$, we take out the maximum weight to
obtain the factor $F$ in front of the state
\begin{equation}
\exp(E_{k}\tau - H\tau)|\mbox{TNS}_{k} \rangle =
F|\mbox{TNS}_{k+1} \rangle ,
\end{equation}
where $|\mbox{TNS}_{k+1} \rangle$ is a normalized TNS. We obtain
the factor $F$ such that $F~*=\lambda_{0}$ whenever any bond is
modified. Because we require no divergence and no convergence to
zero for the state, as in the diffusion Monte Carlo, we adjust the
next energy value $e_{k+1}$ for $F$ to approach 1 in this way:
\begin{equation}
e_{k+1} = e_{k} - \xi \log F ,
\end{equation}
where the value of the feedback parameter $\xi$ is not sensitive
in this algorithm. After we find $e_{k+1}$, we set $F=1$ again for
the next iteration in the computer simulation. We note that during
the time evolution, $e_{k}$ is stable and approaches the
ground-state energy per site in the limit of $k \rightarrow
\infty$. The solution of $|\mbox{TNS}_{\infty} \rangle$ is also
stable.

\section{Numerical Results}

It is instructive to summarize the parameters that are involved in
our task of calculating the ground state of the Hubbard model. The
model Hamiltonian itself contains three parameters. The tensor
network states are defined by the internal-bond dimension, the
spin-bond dimension, and the lattice length. We need the Trotter
parameter, the feedback parameter for energy adjustment, and the
seed for the random number generator we used to set the tensors
and vectors for an initial state in TEBD. Thus, we should set nine
values initially in the simulation:
\begin{eqnarray*}
t, ~~U, ~~\mu             &~~\mbox{in}~~H,\\
\chi, ~~\kappa, ~~L       &~~\mbox{in}~~|\mbox{TNS}\rangle,\\
\tau, ~~\xi,~~\mbox{seed} &~~\mbox{in}~~\mbox{TEBD}.
\end{eqnarray*}
There are several alternative methods for creating initial states;
for instance, all components of tensors and vectors are fixed
intentionally without using the random number generator. In this
case, we need no seed.

Our goal is to find the stable $e$ and $|\mbox{TNS}\rangle$. From
the Suzuki-Trotter decomposition of Eq. (3), we describe the
procedure for the computational simulation:

\begin{enumerate}
\item For a given seed number, all $2\kappa\chi^{4}$
components of the $2L^{2}$ tensors are given by random numbers
between $-0.5$ and $0.5$, and all components of the $5L^{2}$
vectors on the bonds are given by random numbers between $0$ and
$1$. Another option is that, with no seed numbers, all components
of all of the tensors and vectors are given by $1$. After choosing
an initial tensor network state, let $e=0$ and $F=1$.

\item Update the horizontal bonds, the vertical bonds, and then the horizontal bonds in the spin-up layer.

\item Update the horizontal bonds, the vertical bonds, and then the horizontal bonds in the spin-down layer.

\item Update the spin bonds.

\item Update $e$, and set $F=1$. Repeat from step (ii) until $F$ remains stably $1$.
\end{enumerate}

First, we present the typical behavior of the converging energy
$e$ in Fig. 3. We find that, regardless of which nine parameters
are used in our calculations, we obtain similar behavior for $e$
to what is shown in Fig. 3 for all of the other cases. We find
that $\xi$ has no effect on the converging energy value as long as
it is small enough. Furthermore, we have varied the Trotter
parameter $\tau$, and we find that there are no significant
variations in the converging energy $e$ as a function of $\tau$.
Hence, in the main simulations, we fix $\xi=0.03$ and $\tau =
0.02$ for TEBD.

\begin{figure*}
\includegraphics[width= 8.0 cm]{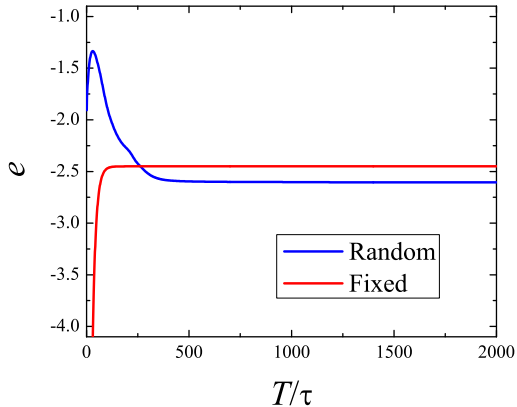}
\caption {The energy value $e$ as a function of the time $T/\tau$
in the two cases of the initial random tensors and the initial
fixed tensors whose components are all 1 for the system of $t=1$,
$U=4$, $\mu=1$, $\chi=2$, $\kappa=2$, $L=10$, $\tau = 0.02$, and
$\xi=0.03$. We let the initial value of $e$ be zero, and we find
that $e$ remains unchanged after $T/\tau=1774$ for
$\mbox{seed}=7733$ and $T/\tau=255$ for no seed. Thus, we obtain
the ground-state energy per site $e=-2.60549$ for the initial
random tensors, and $e=-2.45054$ for the initial fixed tensors.}
\label{fig:fig3}
\end{figure*}

Because it is reasonable that the converging energy value is
independent of any initial state, we should obtain the same
ground-state energy up to the Suzuki-Trotter uncertainty
$\tau^{2}$ as long as $t$, $U$ and $\mu$ are fixed. However, it
seems that there are some barriers in the Hilbert space that
prevent the evolving state from accessing the true ground state.
In other words, if the initial state begins from a topologically
different sector, it will never approach the true ground state in
the process of TEBD. For example, in Fig. 3, we find the
difference between the two converging energy values for the
initial random tensors and the initial tensors whose components
are fixed as 1. Thus, in further calculations, we should repeat
simulations with several seed numbers to study the ground-state
degeneracy and the disjoint space of tensor network states.

By changing the other parameters $t$, $U$, $\mu$, $\chi$,
$\kappa$, and $L$, we can further verify the consistency. It is
obvious that the ground-state energy $e$ should become twice as
large when we simultaneously double $t$, $U$, and $\mu$. We have
checked this consistency so that we can fix the value of $t$ as
usual as $1$. Because the exact ground-state energy for the
non-interacting infinite system \cite{Valenti} is known as
$-1.6211$, we can compare the exact value to our value of
$-1.3422$ for $\chi=\kappa=2$ in the system of $t=1$, $U=0$,
$\mu=0$, and $L=10$. Furthermore, there is another exact result of
the ground-state energy $-1.8514$ at $t=1$, $U=4$, $\mu=0$, and
$L=4$ \cite{Parola}. We compare it to our result of $-1.6508$ for
$\chi=\kappa=2$ and $-1.6539$ for $\chi=\kappa=3$. Because of the
finite values of $\chi$ and $\kappa$, there are some differences
between the exact and ours. We find a tendency for our value to
more closely approach the exact value as we increase $\chi$.
However, the difference of $0.2$ is not small, and increasing
$\chi$ may not improve the ground-state energy significantly. It
indicates that the tensor network state in Fig. 2 may be
incorrect.

In order to find the finite size effect related to $L$, we
calculate the ground-state energy for the Hamiltonian of $t=1$,
$U=4$ and $\mu=0$ by changing $L$. To save computing time, we
perform the calculation for only the easy case of $\chi=\kappa=2$.
We summarize the numerical results for various values of $L$ in
Fig. 4 where we observe the saturation at large $L$.

\begin{figure*}
\includegraphics[width= 8.0 cm]{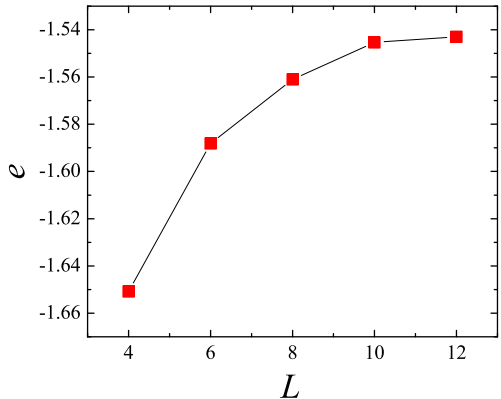}
\caption {The plot of $e$ versus $L$ for the system of $t=1$,
$U=4$, $\mu=0$, $\chi=2$, $\kappa=2$, $\tau = 0.02$, and
$\xi=0.03$. We note the saturation in the limit of $L\rightarrow
\infty$.} \label{fig:fig4}
\end{figure*}

Because the spin-flip symmetry preserving states in Eq. (6) are
living in the subset of the Hilbert space for the spin-flip
symmetry breaking states in Eq. (5), the converging energy for the
state of Eq. (6) should be greater than or equal to the energy for
the state of Eq. (5). For the spin-flip symmetry preserving
states, in the process of TEBD, we perform updating the tensors
and vectors in the spin-up layer, and then we duplicate the
tensors and vectors in the spin-down layer from those in the
spin-up layer. Because we duplicate the bonds in the spin-down
layer, we should modify the factor $F$ such as
$F~*=\lambda_{0}^{2}$ in the process of the horizontal and
vertical bonds updating. We present the numerical results at $\mu
= 0$ in Fig. 5, comparing the ground-state energy of the spin-flip
symmetry preserving state with that of the spin-flip symmetry
breaking state. We note that the ground-state energy of the
spin-flip symmetry preserving state is slightly lower than that of
the spin-flip symmetry breaking state at small $U$. This is caused
by numerical uncertainties, and it is understood as equality. This
means that the symmetry breaking does not take place yet. We find
from Fig. 5 that there is a transition at $U=0.39(1)$ for $\mu=0$.
At large $U$, the ground-state energy of the spin-flip symmetry
breaking state depends heavily on $U$. We note that the
ground-state energy of the symmetry preserving state is almost
independent of $U$. This independence means that the expectation
value of the number operator is given by $\langle n_{i\uparrow}
\rangle = \langle n_{i\downarrow} \rangle \approx \frac{1}{2}$ for
any $i$.


We conclude that our method is effective in searching for the
ground state of the Hubbard model. We emphasize that it is
possible to determine the energy and the ground state for any
chemical potential.

\begin{figure*}
\includegraphics[width= 8.0 cm]{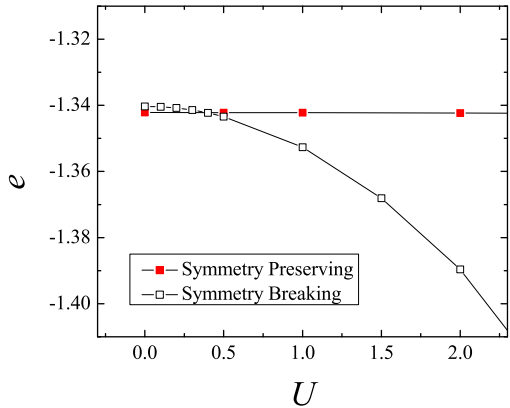}
\caption {The ground-state energy $e$ versus $U$ for the system of
$t=1$, $\mu = 0$, $\chi=2$, $\kappa=2$, $L=10$, $\tau=0.02$, and
$\xi=0.03$ with respect to the spin-flip symmetry breaking state
of Eq. (5) and the spin-flip symmetry preserving state of Eq. (6).
At small $U$, the ground-state energy for the state of Eq. (6) is
roughly the same as that for the state of Eq. (5). We find a
cut-point at $U=0.39(1)$. This is a signal of phase transition
that takes place at $U=0.39(1)$ for $\mu=0$. For large $U$, the
spin-flip symmetry breaking state is the true ground state in the
model. } \label{fig:fig5}
\end{figure*}

\section{Conclusion}

In summary, we have presented a method for obtaining the
ground-state energy and the wave function for two-dimensional
quantum many-fermion systems, especially the Hubbard model. We may
call this method diffusive TEBD. Because there is a certain
discrepancy between the exact ground-state energy and our value
for the TNS, it is still questionable whether or not the TNS is
correct and the diffusive TEBD is useful. We suggest that the
diffusive TEBD is an effective method.

Although we built a user-friendly library in the framework of
previous computer code \cite{Chung2}, we obtain only preliminary
numerical results because we use the full SVD, which is very
inefficient. In future work, we will implement an SVD package
based on the Lanczos algorithm with partial reorthogonalization
\cite{Simon} to find only a few eigenvectors and their
corresponding singular values, which are sufficient for our
truncation scheme.

When we use multi-core computers, it is possible to parallelize
the local updates of the horizontal bonds and the spin bonds. For
the vertical-bond update, we may apply the concept of a pipeline
to optimize the roles of the multiple cores. We anticipate
progress in this parallel computing scheme.

In future work, for a fixed $U$, we need to investigate whether
any phase transitions happen as we change the chemical potential
$\mu$ in the Hamiltonian. If there are any transitions in
simulations, the phase transitions may be related to topological
orders \cite{Wen} or the topological entanglement entropy
\cite{Kitaev}. In connection with topological orders, we should
give a definitive answer to the ground-state degeneracies.
Furthermore, it is necessary to perform the same simulation by
changing periodic or open boundary conditions.

It is of interest to extend our method to the case of two-body
interactions. A typical topic of interest for two-body
interactions may be the fractional quantum Hall effect, for which
MPS can be used as an accessible subset of the huge Hilbert space.
In the fractional quantum Hall effect, the energy gap between the
ground state and the first excited state provides a lesser
entanglement entropy, which makes it possible to use MPS with a
relatively small bond dimension.

\section*{Acknowledgments}
This work was partially supported by the Basic Science Research
Program through the National Research Foundation of Korea (NRF)
funded by the Ministry of Education, Science and Technology (Grant
No. 2011-0023395) and by the Supercomputing Center at Korea
Institute of Science and Technology Information with their
supercomputing resources, including technical support (Grant No.
KSC-2012-C1-09). The author would like to thank Michelle Ebbs for
reading the manuscript.

\end{document}